# Optical Vortex Shaping & Multiple Singularities Manipulation via High-order Cross-phase

CHEN WANG[1] YUAN REN,[1,2,*] TONG LIU,[1] SONG QIU,[1] ZHIMENG LI,[1] AND HAO WU[1]

[1]*Department of Aerospace Science and Technology, Space Engineering University, Beijing 101416, China*
[2]*State Key Lab of Laser Propulsion & its Application, Space Engineering University, Beijing 101416, China*
*[*renyuan_823@aliyun.com](*renyuan_823@aliyun.com)*

**Abstract:** Increasing demand for practical applications is forcing deeper research into optical vortices (OVs): from the generation and measurement to shaping and multiple singularities manipulation of OVs. Herein, we propose a new type of phase structure called the high-order cross phase (HOCP) can be employed to modulate OVs to implement both polygonal shaping and singularities manipulation. Theoretically, we investigate the propagation characteristics of OVs with the HOCP. In experiments, we achieve the shaping and singularities manipulation of OVs by utilizing the HOCP. On this basis, we discuss the interference patterns of superposed OVs after the modulation. This work provides an alternative method to achieve both polygonal shaping and multiple singularities manipulation, which will facilitate applications in optical micro-manipulation, optical communication, and high-dimensional quantum entanglement.

## 1. Introduction

Optical vortices (OVs) with the phase factor $\exp(im\phi)$, carry orbital angular momentum (OAM) of $m\hbar$ per photon, where $m$ denotes the topological charges (TCs), $\phi$ denotes the azimuthal angle and $\hbar$ is the Planck constant. Nowadays, OVs are motivating plethora of applications in the field of optical micro-manipulation[1], high-dimensional quantum entanglement[2] and remote sensing of the angular rotation of structured objects via the optical rotation doppler effect[3, 4], etc. In turn, increasing demand for practical applications is forcing deeper research into OVs: from generation[5] and measurement[6] to shaping[7-9] and singularities manipulation[10-12] of OVs.

    Recently, the cross phase (CP), a new type of phase structure has been employed to achieve mutual conversion between Laguerre-Gauss (LG) beams and Hermite-Gaussian beams (HG)[13] that opens up a new horizon for generation and measurement of high-order OVs[14], which has the form

$$\psi_0'(x, y) = u(x\cos\theta - y\sin\theta)(x\sin\theta + y\cos\theta) \tag{1}$$

where $(x, y)$ denotes Cartesian coordinates, the coefficient $u$ controls the conversion rate, and the azimuth factor $\theta$ characterizes the rotation angle of converted beams in one certain plane. The method with the CP and the method of the mode conversion are both based on the astigmatism principle, while the former avoids the use of optical elements such as cylindrical lenses but holograms instead, which is more conducive to the precise manipulation of the light field and greatly evades the harsh requirements of relative position for cylindrical lenses. It is noteworthy that Eqs.(1) could be simplified to $\psi'(x, y) = uxy$ when $\theta = 0$ and we only take this typical situation into count in this paper.

In this paper, inspired by the CP, we propose a new type of phase structure, the *high-order cross-phase* (HOCP), which can be employed to achieve both polygonal shaping and singularities manipulation. The shape of the polygon can be modulated by the order of the HOCP and TCs, and the singularities distributions can be controlled by the parameter $u$. The HOCP could modulate the shape and the singularities distribution of OVs and would not convert OVs to HG beams like the CP, despite OVs with the HOCP are also a kind of non-eigenmode. An OV carrying the HOCP, which can be regarded as a new type of light field, otherwise, can be considered that the HOCP is attached to the OV to modulate the light field further. We would like to choose the latter view.

## 2. Theory

May the HOCP

$$\psi(x, y) = u x^p y^q \tag{2}$$

where $p$ and $q$ are positive integer exponents and the sum of two (the order of the phase) is 3 and above, namely, the $3^{rd}$, the $4^{th}$, the $5^{th}$ CP, etc. For simplicity and without loss of generality, this paper focuses on the analysis of the $3^{rd}$ and the $4^{th}$ CP. Meanwhile, If the sum equals to 2, the HOCP here degrades into the CP, i.e. the $2^{nd}$ CP, which we call low-order CP (LO-CP) hereinafter.

### A. The distributions of the HOCP

The phase distribution of the LO-CP (the $2^{nd}$ CP) and the HOCP (the $3^{rd}$ CP and $4^{th}$ CP) are shown in Fig. 1(a). In the slowly varying part of the center, the phase distribution and symmetry axes vary by orders. However, the HOCP has the hyperbolic distributions similar to the LO-CP in sharply varying parts. That's why we still call it CP, albeit higher-order. Fig. 1(b) shows the phase distributions of $LG_{05}$ with corresponding phase in Fig. 1(a), and the red dotted lines characterize the singularities distributions, which indicated the TCs of LG beams employed here are 5. It's worthy note that the values of the parameter $u$ we set here are dramatical high. Whether in simulations or the design of holograms, we consider the real situations here rather than nondimensionalize of ideal situations.

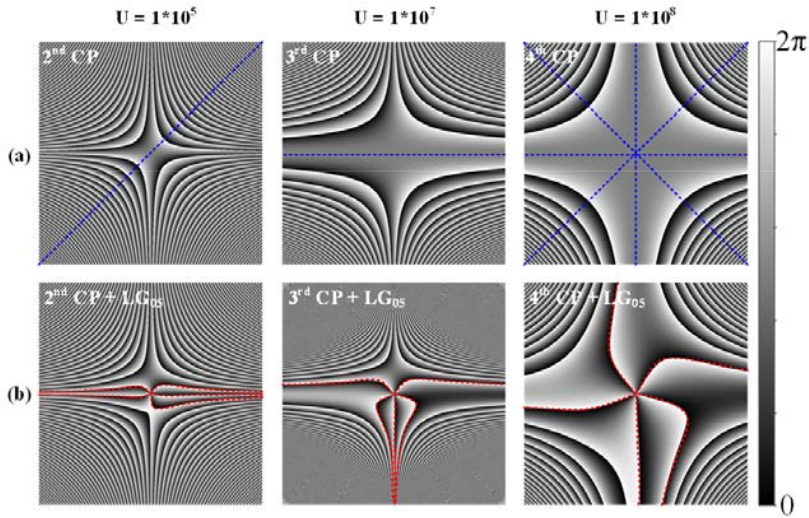

Fig. 1 The phase distributions of LG beams with the LO-CP and the HOCP. (a) The phase distributions of the LO-CP (the $2^{nd}$ CP) and the HOCP (the $3^{rd}$ and the $4^{th}$ CP), and the blue dotted lines are symmetry axes. (b) The phase distributions of $LG_{05}$ with corresponding phase in (a), and the red dotted lines characterize the singularities distributions. The parameter $u$ in the $2^{nd}$ CP, the $3^{rd}$, and the $4^{th}$ CP are $1*10^5$, $1*10^7$, $1*10^8$, respectively.

## B. Propagation characteristics

For fully demonstrating the modulation process of the HOCP, we simulate the propagation of OVs (Take $LG_{03}$ as an example) modulated by the HOCP (the $3^{rd}$ and $4^{th}$ CP) as shown in Fig. 2 and Fig. 3. Fig. 2(a) depicts the phase distributions. With the increase of the propagation distance, the phase distributions gradually become the phase of OV with TC of 3, but the singularities at the center are split. By utilizing the $3^{rd}$ CP, the intensity distributions of $LG_{03}$ change from a ring to a triangle as shown in Fig. 2(b). In addition, in order to show the application value of the HOCP in optical micro-manipulation and optical communication, we also investigate the OAM spectrums[15] and density distributions[16] as shown in Fig. 2(c) and Fig. 2(d). The research shows that the distributions of OAM densities are exactly same as the intensity distributions, and the HOCP keep the OAM spectrums of original OVs. During the modulation process, the mode purities have hardly changed, and they have remained above 99%.

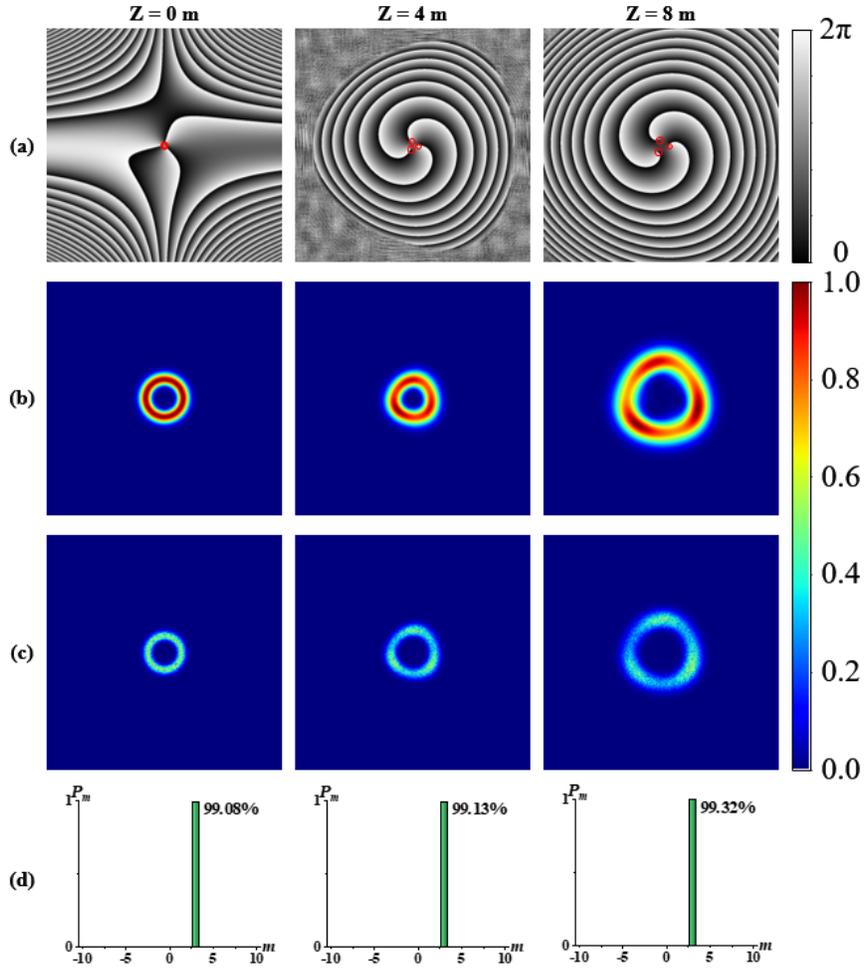

Fig. 2 The simulated propagation of $LG_{03}$ with the $3^{rd}$ CP at different distances. (a) The phase distributions, and the red circles indicates the position of the singularities. (b) The intensity distributions. (c) The density distributions of OAM. (d) The OAM spectrums, and the numbers marked on the side of the bars are purities of the corresponding mode.

Akin to the Fig. 2, the 4$^{th}$ CP shares the similar characteristics as shown in Fig. 3. It's to be noted that the 4$^{th}$ CP only shapes the distribution of LG$_{03}$ to square, and the TCs are still equal to 3.

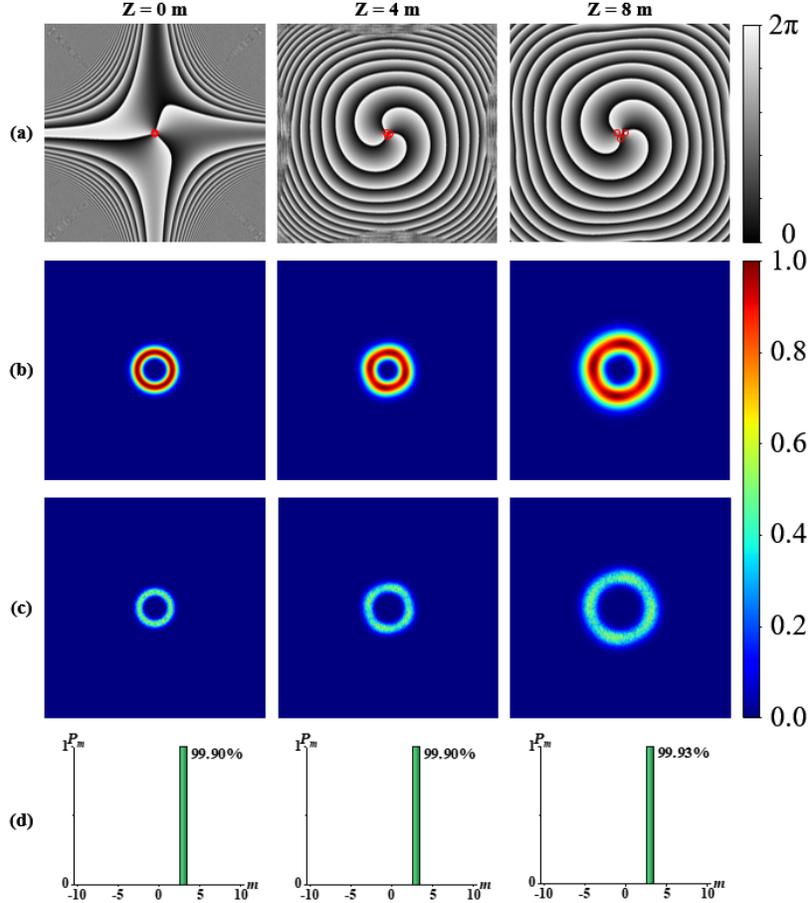

Fig. 3 The simulated propagation of LG$_{03}$ with the 4$^{th}$ CP at different distances. (a) The phase distributions, and the red circles indicates the position of the singularities. (b) The intensity distributions. (c) The density distributions of OAM. (d) The OAM spectrums, and the numbers marked on the side of the bars are purities of the corresponding mode.

It's worthy of note that for keeping the low conversion rate to fully demonstrate the conversion process despite the large transmission distance, the values of parameters $u$ adopted in Fig. 2 and Fig. 3 are relative low as $1*10^8$ and $1*10^{10}$. Moreover, the modulation results hardly change in far-field conditions. Therefore, it is not necessary to recalculate the propagation longer than 8m.

## 3. Experiments and results

The experimental setup for generation and measurement of high-order OVs by utilizing the CP is shown in Fig. 4. The laser delivers a collimated Gaussian beam with wavelengths of 632.8nm after a linear polarizer (LP), a half-wave plate (HWP) and a telescope consists of two lenses (L1, L2) are used for collimation. The combination of the LP and the HWP is served to rotate the laser polarization state along the long display axis of SLM (by setting the LP polarization in the vertical direction) and adjust the power of an incident light on SLM (by rotating the HWP). The SLM (HOLOEYE PLUTO-NIR-011) precisely modulate the incident light via loading a hologram and then the aperture (AP) is used to select the first diffraction order of the beam to avoid other stray light. The beam is adjusted to an appropriate size through another telescope (L3, L4). Finally, L5 changes the incident light to far field and the intensity pattern is registered by a CCD camera (NEWPORT LBP2) placed at the back focal plane of L5.

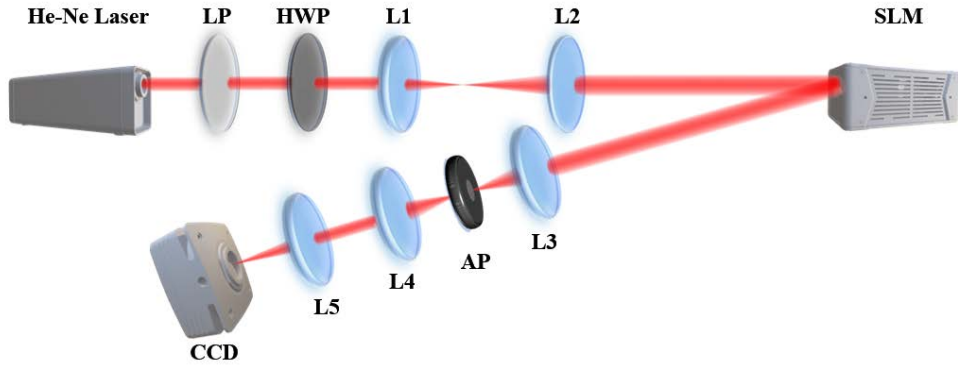

Fig. 4. Experimental setup for generation and measurement of high-order OVs by utilizing the CP. LP: linear polarizer. HWP: half-wave plate. L: lens. SLM: spatial light modulator. AP: aperture. CCD: charge couple device.

## A. Singularities manipulation

For simplicity and without loss of generality, we utilize the HOCP (the 3$^{rd}$ CP) to implement the manipulation of singularities of OVs (take LG$_{03}$ for example). By adjusting the value of the parameters $u$ from 0 to $4*10^8$, the intensity distributions are shaped from ring to triangle as shown in Fig. 5(b), which the experimental results of Fig. 5(d) agree well with. Meanwhile, the distance from singularities to center increases as the parameters $u$ increases. Because of the unlimited value of the parameter $u$, which we can alter to adjust the position of singularities precisely.

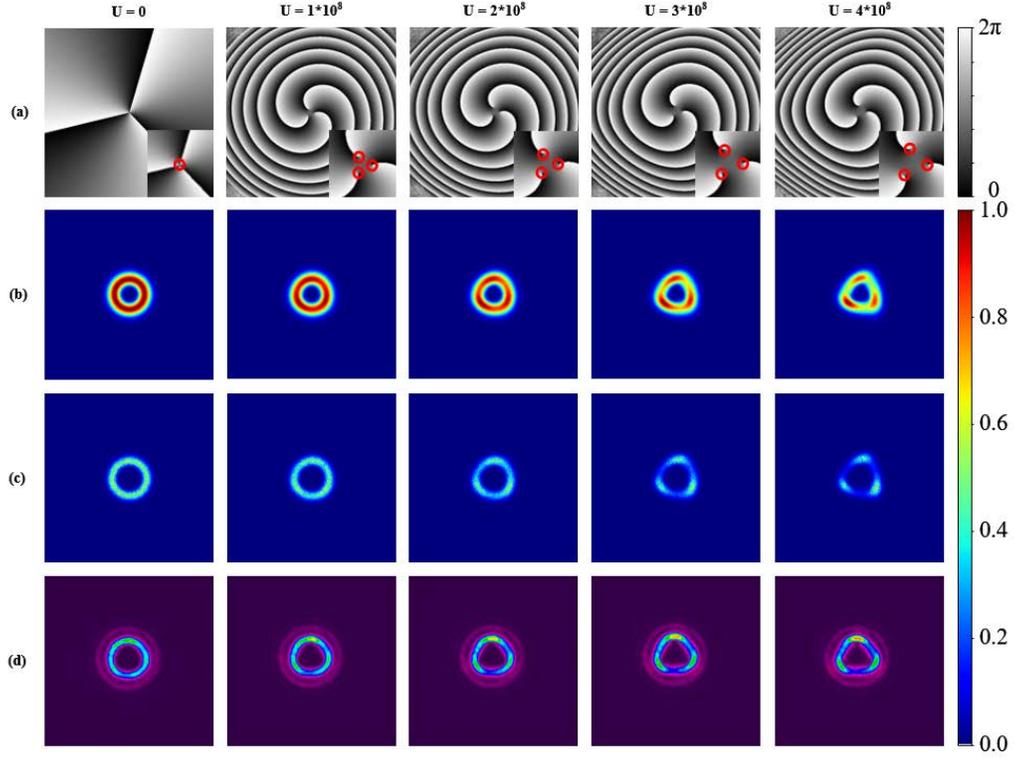

Fig. 5. The singularities manipulation of OVs ($LG_{03}$) by utilizing the HOCP (the $3^{rd}$ CP) at far-field with different parameters $u$. (a) The phase distributions, and the scale of the local phase magnification is 1:10. The red circles indicates the position of the singularities. (b) The simulated intensity distributions. (c) The density distributions of OAM. (d) The experimental intensity distributions.

## B. Polygonal shaping

As shown in Fig. 2 and Fig. 3, we can shape the OVs (take $LG_{03}$ as an example) with HOCP (the $3^{rd}$ and $4^{th}$ CP). Actually, the shape of the OVs after modulation is closely related to the orders of the HOCP, which is equal to the number of sides of the polygon. Fig. 6(b) shows the triangle, quadrilateral, pentagon, and hexagon distributions of OVs where the corresponding orders of the employed HOCP are 3,4,5 and 6, respectively. The phase distributions and OAM density distributions are also calculated, which are depicted in Fig. 6(a) and Fig. 6(c).

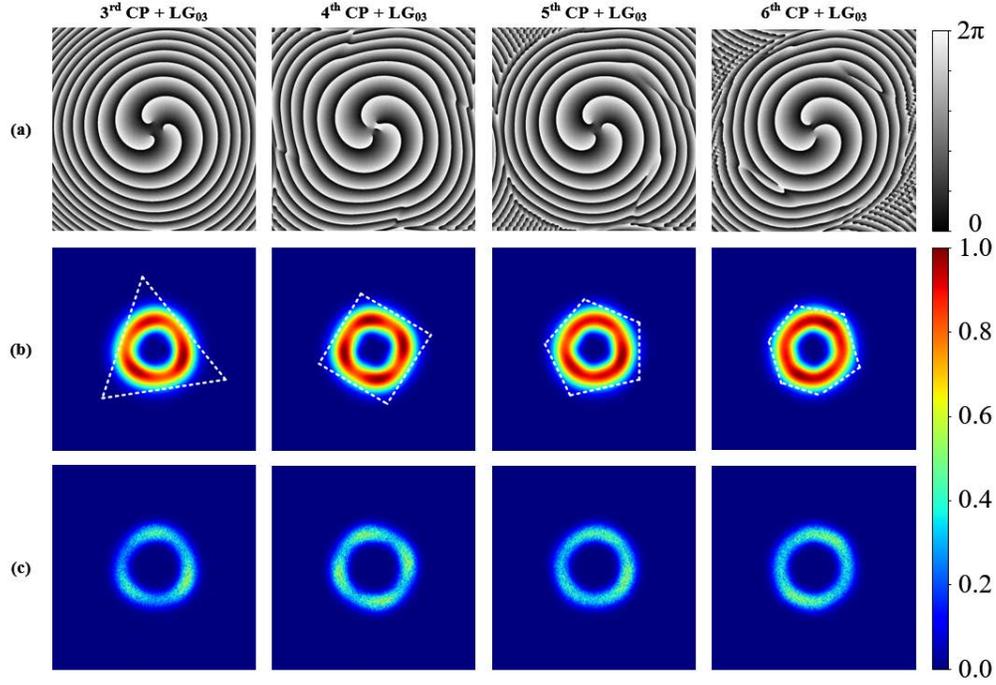

Fig. 6 Polygonal shaping to OVs by utilizing the HOCP. (a) The phase distributions of OVs after shaping. (b) The intensity distributions. (c) The OAM densities.

The size of the light field distribution is also an important part of the shape. We could modulate the size of the polygonal OVs by TCs shown in Fig. 7. We adopt the $4^{th}$ CP to implement the size modulation of the polygonal OVs by changing the TCs as 1,4,7 and 11. The function of TCs is like a magnifying glass: with the increase of TCs, the size of the light field also gradually increases, but the shape of the light field remains a quadrilateral due to the constant $u$ as shown in Fig. 7(b). This simulated intensity distributions agree well with the experimental results shown in Fig. 7(c).

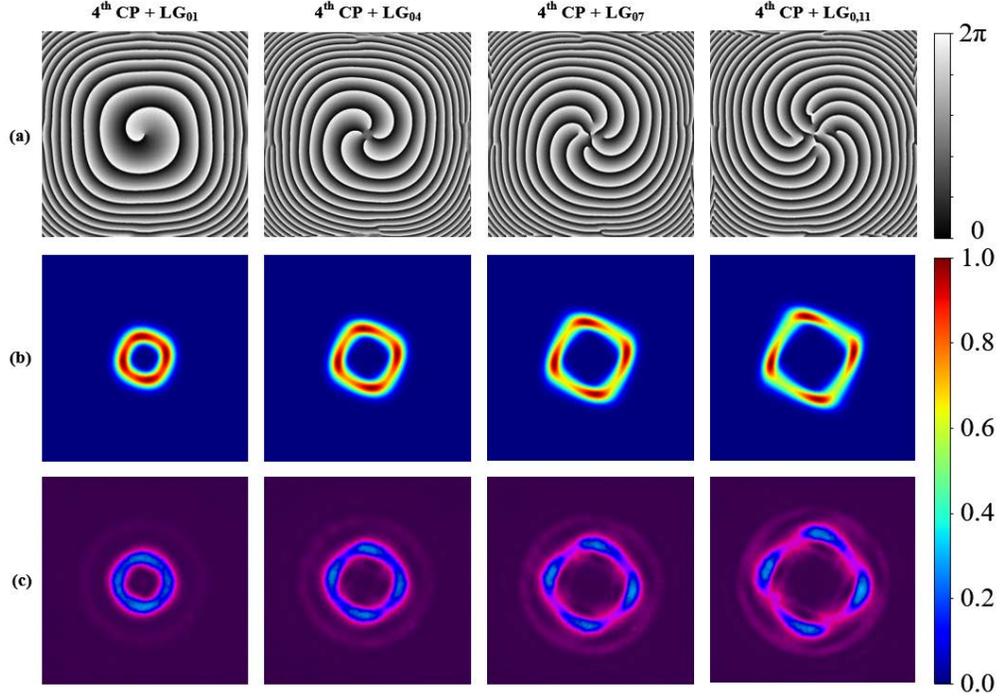

Fig. 7 The modulation of the size of the polygonal OVs by TCs. (a) The phase distributions of OVs after size modulation of the polygonal OVs. (b) The intensity distributions. (c) The OAM densities.

## C. Interference pattern

To further demonstrate that the HOCP only manipulates the singularity distributions, but does not affect the OAM spectrums, we also discuss the situation of superimposed OVs ($LG_{1\pm1}$, $LG_{1\pm2}$, $LG_{1\pm3}$, $LG_{1\pm4}$) with the HOCP (the 4$^{th}$ CP). Fig. 8(a) shows the intensity distributions of the initial OVs, whose superposed intensity distributions are simulated in Fig. 8(c) that agrees well with the experimental results in Fig. 8(d). We can tell from the results that the polygonal OVs still maintain the original interference characteristics as Fig. 8(a). The angular distributions of petals indicate the TCs of this superposed OVs, and the radial distributions of petals indicate the radial nod is 1. Fig. 8(b) shows the simulated phase distributions of the superposed OVs with the HOCP. Fig. 8(e) shows the simulated OAM distributions of the superposed OVs with the HOCP and the results denote the OAM spectrums are same with the normal superposed OVs, which we can tell from that the HOCP does not affect the OAM spectrums.

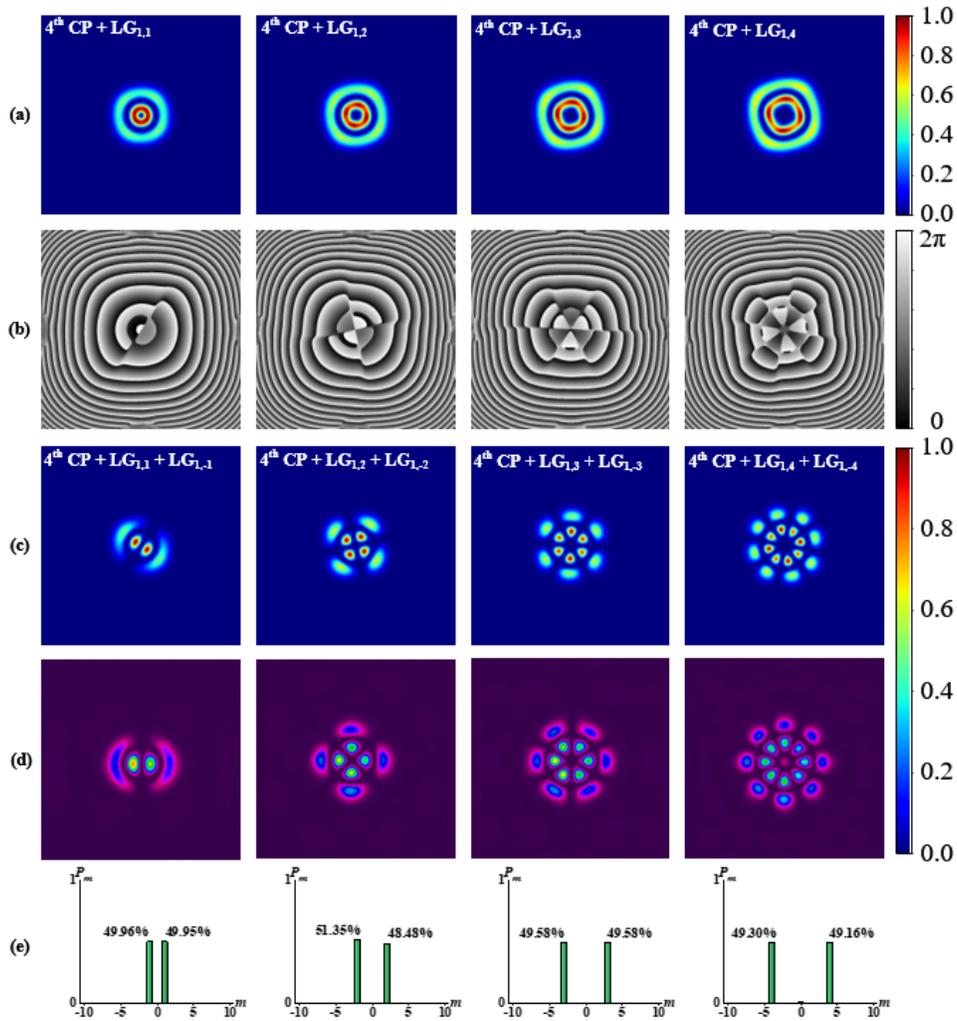

Fig. 8 The Interference patterns of superimposed OVs with the HOCP. (a) The simulated intensity distributions of the initial OVs with the HOCP. (b) The phase distributions of the superimposed OVs with the HOCP. (c) The simulated intensity distributions. (d) The experimental intensity distributions. (e) The OAM spectrums.

**References**

1. Y. Q. Zhang, X. Y. Zeng, L. Ma, R. R. Zhang, Z. J. Zhan, C. Chen, X. R. Ren, C. W. He, C. X. Liu, and C. F. Cheng, "Manipulation for Superposition of Orbital Angular Momentum States in Surface Plasmon Polaritons," Adv. Opt. Mater. **7**, 9 (2019).
2. Y. J. Shen, X. J. Wang, Z. W. Xie, C. J. Min, X. Fu, Q. Liu, M. L. Gong, and X. C. Yuan, "Optical vortices 30 years on: OAM manipulation from topological charge to multiple singularities," Light-Sci. Appl. **8**, 29 (2019).
3. S. Qiu, T. Liu, Z. Li, C. Wang, Y. Ren, Q. Shao, and C. Xing, "Influence of lateral misalignment on the optical rotational Doppler effect," Appl Opt **58**, 2650-2655 (2019).
4. W. Zhang, D. Zhang, X. Qiu, and L. Chen, "Quantum remote sensing of the angular rotation of structured objects," Physical Review A **100**, 043832 (2019).

5.	Y. Yang, Q. Zhao, L. Liu, Y. Liu, C. Rosales-Guzmán, and C.-w. Qiu, "Manipulation of Orbital-Angular-Momentum Spectrum Using Pinhole Plates," Physical Review Applied **12**(2019).
6.	L. Chen, W. Zhang, Q. Lu, and X. Lin, "Making and identifying optical superpositions of high orbital angular momenta," Physical Review A **88**(2013).
7.	Y. Shen, Z. Wan, Y. Meng, X. Fu, and M. Gong, "Polygonal Vortex Beams," IEEE Photonics Journal **10**, 1-16 (2018).
8.	D. Yang, Y. Li, D. Deng, J. Ye, Y. Liu, and J. Lin, "Controllable rotation of multiplexing elliptic optical vortices," Journal of Physics D: Applied Physics **52**(2019).
9.	D. Shen, K. Wang, and D. Zhao, "Generation and propagation of a new kind of power-exponent-phase vortex beam," Optics express **27**(2019).
10.	J. Leach, M. R. Dennis, J. Courtial, and M. J. Padgett, "Vortex knots in light," New Journal of Physics **7**, 55-55 (2005).
11.	T. Yoshitaka and Z. Shukui, "Split in phase singularities of an optical vortex by off-axis diffraction through a simple circular aperture," Optics Letters **42**, 1373-.
12.	B. S. B. Ram, A. Sharma, and P. Senthilkumaran, "Diffraction of V-point singularities through triangular apertures," **25**, 10270-10275 (2017).
13.	G. Liang and Q. Wang, "Controllable conversion between Hermite Gaussian and Laguerre Gaussian modes due to cross phase," Optics express **27**, 10684-10691 (2019).
14.	D. Shen and D. Zhao, "Measuring the topological charge of optical vortices with a twisting phase," Opt Lett **44**, 2334-2337 (2019).
15.	L. Torner, J. Torres, and S. Carrasco, "Digital spiral imaging," Optics express **13**, 873-881 (2005).
16.	A. Dudley, I. A. Litvin, and A. Forbes, "Quantitative measurement of the orbital angular momentum density of light," Appl Opt **51**, 823-833 (2012).